\documentclass[preprint2]{aastex}
\slugcomment{submitted to Astrophysical Journal Letters}
\shorttitle{Weak Lensing in a Nearby Cluster}
\shortauthors{Joffre, et al.}

\begin{document}


\title{Weak Gravitational Lensing by the Nearby Cluster Abell 3667}


\author{Michael Joffre\altaffilmark{1,2,7}, Philippe
Fischer\altaffilmark{3,7}, Joshua Frieman\altaffilmark{1,2,7}, Timothy McKay\altaffilmark{4,7}, Joseph J. 
Mohr\altaffilmark{1,6,7}, Robert C. Nichol\altaffilmark{5,7}, David
Johnston\altaffilmark{1,2}, Erin
Sheldon\altaffilmark{4} and Gary Bernstein\altaffilmark{3}}
\altaffiltext{1}{Dept. of Astronomy and Astrophysics, University of Chicago, Chicago
IL \\ Email:joffre@coltrane.uchicago.edu}
\altaffiltext{2}{NASA/Fermilab Astrophysics Center, Fermi National Accelerator Lab,
Batavia, IL}
\altaffiltext{3}{Dept. of Astronomy, University of Michigan, Ann Arbor, MI}
\altaffiltext{4}{Dept. of Physics, University of Michigan, Ann Arbor, MI}
\altaffiltext{5}{Physics Department, Carnegie Mellon University, Pittsburgh, PA}
\altaffiltext{6}{Chandra Fellow}
\altaffiltext{7}{Visiting Astronomer, National Optical Astronomy Observatories, which is 
operated by the Association of Universities for Research in Astronomy, Inc., under contract to 
the National Science Foundation}

\begin{abstract}
We present two weak lensing reconstructions of the nearby ($z_{cl}=0.055$) 
merging cluster Abell 3667, based on observations 
taken $\sim 1$ year apart under different seeing conditions. This is the 
lowest 
redshift cluster with a weak lensing mass reconstruction to date. 
The reproducibility of features in the two mass maps demonstrate that weak lensing studies of low-redshift clusters are feasible. These data constitute the 
first results from an X-ray luminosity-selected weak lensing survey of 19 low-redshift ($z< 0.1$) southern clusters.

\end{abstract}



\newpage
\section{Introduction}

Weak gravitational lensing---measurement of the induced shear of distant 
galaxy images to infer the foreground mass distribution---has now become 
a standard tool to probe dark matter in galaxy clusters. 
Since the work of Tyson et al. (1990), 
weak lensing studies of over twenty galaxy clusters have been published 
\citep{mellier99}. In all cases to date, the observations have been confined  
to moderate to high-redshift clusters, $z_{cl} >
0.15$\renewcommand{\thefootnote}{\fnsymbol{footnote}}\footnote[1]{Lensed arcs have been detected in two low-redshift
\renewcommand{\thefootnote}{\arabic{footnote}} 
clusters \citep{bm98,cam98}, but such instances of strong lensing only 
probe the cores of clusters.}, for two reasons:
(i) the cluster lensing strength is maximized if the 
angular diameter distance to the foreground cluster is roughly half that of the
background source galaxies, which at a limiting depth of $R \sim 25$ are typically at $\langle z_s \rangle \sim 1$; 
(ii) a distant cluster subtends an area on the sky which can be roughly 
encompassed 
by a single CCD chip, allowing deep exposures of the entire field in 
moderate observing time. 
Unfortunately, most of our knowledge of cluster properties comes from 
X-ray and dynamical (optical redshift) studies, which are much more easily conducted for nearby clusters. This
mismatch between the observing requirements for weak lensing and dynamical
studies has hampered direct comparison of these mass measuring techniques.  It
would be beneficial to have weak lensing observations for a sample of nearby
clusters. This  
would enable more detailed
studies of the cluster binding mass, baryon fraction, and morphology. 

For weak lensing studies, nearby clusters have the advantage that their
background galaxies are relatively well resolved (reducing the effect of PSF
smearing) and their inferred projected mass distributions are relatively
insensitive to the uncertain background source galaxy redshift distribution in the limit $z_{cl} \ll
z_s$. The advent of CCD mosaic cameras on 4m telescopes, coupled with developments in weak lensing analysis, now make possible wide-field studies of weak lensing 
in nearby clusters \citep{jof99, ste96}. 

With these benefits in mind, we have 
begun a complete, X-ray luminosity-selected weak lensing survey of 19
nearby ($z_{cl} < 0.1$) Southern clusters \citep{jof99}; this sample, drawn from the
XBACS \citep{ebe96}, is also 
being targeted by the Viper telescope for Sunyaev-Zel'dovich observations 
\citep{rom99}. As of October 1999, we have completed imaging observations 
of 9 clusters in the sample, using the BTC and Mosaic II cameras at the 4m 
telescope at CTIO. In this Letter, we present results for our first target, the $z=0.055$ cluster Abell 3667.

The major obstacle to weak lensing studies of nearby clusters is the
small shear signal they produce. Shear estimates are limited by the intrinsic
ellipticity of the background galaxies (shape noise) and PSF anisotropies.  To create reliable mass maps we must therefore 
measure ellipticities for a large number of background galaxies and 
correct for the PSF anisotropy very accurately. Given that we expect cluster-induced shears of at most $\sim 5\%$, we have reduced systematic sources of shear, which include atmospheric turbulence, 
telescope shake, and optical aberration, to
less than $0.5\%$ across the images (based on the corrected stellar ellipticities). As proof of our ability to make these
measurements, we present weak lensing analyses of A3667 based on two sets of 
observations taken $\sim 1$ year 
apart under different seeing conditions and with different 
observing strategies.

\section{Observations}

Much ancillary information is available for Abell 3667. Based on 
redshifts for 154 cluster members 
\citep{sod92,kat96}, an optical velocity dispersion in the range 
$\sigma_v = 970 - 1200$ km/s has been inferred~\citep{sod92,fad96}.  
From the ROSAT All-Sky Survey, the  X-ray luminosity  $L_X = 
6.5 \times 10^{44}$
ergs/s \citep{ebe96}, among the strongest X-ray sources in the 
southern sky. The radio map shows a double halo structure (Rottgering et al.~1997).  
The A3667 field contains 
two dominant D galaxies; the ROSAT PSPC image \citep{kno96} and ASCA 
temperature map \citep{mar98} indicate 
the cluster is undergoing a merger along the
direction \citep{kno96} joining them.

We observed A3667 on two separate runs with 
the BTC \citep{wit98} at the CTIO 4m telescope.  The BTC is a $
4096\times 4096 $ pixel camera with a pixel scale of $0.43 ''$; there are 
significant
gaps between the four chips that must be removed from the final combined 
image by dithering exposures. In June 1997, two of us 
(PF and JM) observed A3667 in the $R$ and $B_J$ bands in
relatively poor seeing (combined 
stellar FWHM of $1.55'' \pm 0.15''$ in R). 
The combined image in each filter covers an approximate area 
$42' \times 42'$, with a maximum surface brightness limit of  28.6 
mag arcsec$^{-2}$ in $R$ and
28.9 mag arcsec$^{-2}$ in
$B_J$, corresponding to $1 \sigma$ in the sky.  Total observing times were 5500s in $B_J$ and 15500s in $R$. We denote these 
the `$\alpha$' set of images in our analysis. In September 1998, 
three of us (JF, TM, RN) 
observed A3667 under better seeing conditions (combined stellar FWHM
$ = 1.23'' \pm 0.07''$ in $R$), covering
a $44' \times 44'$ area to a maximum surface brightness
limit of 28.2 mag arcsec$^{-2}$ in $R$ and 27.9 mag arcsec$^{-2}$ in $B$. This
corresponded to a total observing time of 12600s in $R$ and 2250s in $B$.  
We also obtained $I$-band images which we have used to construct 
a color-magnitude diagram to 
remove cluster members from the background sample. (Due to fringing 
effects, the $I$-band data was not used in the lensing analysis itself.)
We denote these better-seeing images the `$\beta$' set.

After debiasing and flatfielding of the frames, the frames were coadded. BTC 
observations of 
USNO astrometric fields were used to remove field distortion from the images. Objects were 
detected using SExtractor (v2.1.0) \citep{ber96}. Our own code measures quadrupole 
moments $Q_{ij}$ of each object using the 
adaptive Gaussian weighting scheme of Bernstein, et al.(2000): 
the image is multiplied by an elliptical Gaussian weight function, 
the shape of which is successively iterated 
to match the object's ellipticity. This routine returns estimates of the   
ellipticity vectors $e_1 =(Q_{11}-Q_{22})/(Q_{11}+Q_{22})$ and $e_2= 2
Q_{12}/(Q_{11}+Q_{22})$ and their uncertainties. The unsaturated 
bright stars in each image were used to characterize the PSF anisotropy 
as a function of position. Following Fischer \& Tyson (1997), 
we convolved the images with a spatially varying kernel which 
circularizes the stellar images. We then repeat the detection and 
measurement of background galaxies in the PSF-corrected images. 
The moments of each galaxy are finally corrected for isotropic 
PSF dilution using the simulations and analytic results of 
Bernstein, et al.(2000); the correction factor depends upon galaxy 
image size relative to the PSF and upon the image profile. 
Objects in the background galaxy samples for the lensing analysis 
are selected to have magnitudes $R<24.75 $, $B<24.5 $, and $B_J<25.5 $. They are also required to 
have half-light radii at least $1.5$ times that of the PSF.  The $R-I$ color-magnitude 
diagram is used to remove red cluster members from the 
background sample brighter than $R =22$. These cuts ensure that
the object moments can be accurately measured ($S/N > 8-10 \sigma$), that stellar
contamination is minimal, and that the vast majority of the sample lies well behind
the cluster. The resulting samples 
contain approximately 11,000 background galaxies for the 
$\alpha$ R image, 30,000 for the $\beta$ R, 18,000 for the $\alpha$ 
$B_J$, and 11,000
objects for the $\beta$ $B$ image.  We note that the $\beta$ R image has a larger
number of galaxies than the $\alpha$ set, although the later has a longer exposure time.  This is due to the fact that the $\beta$
images covered a larger area on the sky and the poor seeing of the
$\alpha$ R image makes galaxies and stars difficult to distinguish.  Many of these
smaller galaxies are lost when we cut on half-light radius.

\section{Comparison}
To study the robustness of the mass maps derived from the two sets of 
observations, we first examine 
 the consistency of the measured ellipticities of background galaxies.  We
focus on the shear measurements (as opposed to the mass maps themselves) 
because we expect the errors to be uncorrelated in
different regions of the sky.  In principle, differences in 
seeing, imaging
depth, and filters will lead to differences between the shear 
fields estimated from the two observations. However, after applying 
the corrections discussed above, the derived shear fields should be 
strongly correlated.

We trimmed the R and B band 
images in the $\alpha$ and $\beta$ datasets to the 
regions common to both 
and calculated the mean ellipticities 
of the background galaxies 
in $100$ angular bins of width 
$\sim 25''$. To quantify the consistency of the two fields we
calculate the $\chi^2$ value for the ellipticities for each filter,
$$\chi^2 = \sum_{i=1}^2 \sum_{N_b=1}^{100} \frac{(e_{i\alpha}-e_{i\beta})^2
}{(\sigma^2_{e_{i\alpha}}+\sigma^2_{e_{i\beta}})}~.$$ 
The estimate of the ellipticity variance in a spatial bin for a given 
data set is
$$\sigma_{ei}^2= (N_c/N^2) \langle
\sigma_m^2 \rangle_{c} +(N_d/N)^2 \sigma_{rms}^2 /(N_d-1),$$ where $N$ is
the total number of background galaxies in the bin and $N_c$ and $N_d$ are the numbers
of galaxies in the bin with and without counterparts in the other data set.
$\langle \sigma_m^2 \rangle_c$ is the average measurement error in $e_i$ derived from our
measurement uncertainties in $Q_{ij}$ for the galaxies common to both sets; 
$\sigma_{rms}^{}$ is the {\it rms} spread in the ellipticities of the $N_d$ galaxies in the bin not
found in the other set. That is, for galaxies measured in both data sets, 
we use the measurement uncertainty rather than the rms per bin to take 
into account the correlation between the data sets. 

Figure 1a shows the binned 
ellipticities of background galaxies found in both data  
sets in the 
R band; the reduced $\chi^2$ is 1.11 for 200 d.o.f. 
Fig. 1b compares the ellipticities for all galaxies: 
in this case, the scatter between the datasets 
is significantly larger, as expected since they are no longer confined to the
same galaxies;
the reduced 
$\chi^2$ for all the  objects is
1.19, implying a probability 
$P(\chi^2|200) = 3.5\%$ in the case of random errors. The average ellipticity in Fig. 1b is $2.57 \pm 0.22\%$ for
the $\alpha$ set and $2.42 \pm 0.16\%$ for the $\beta$ 
set.  As expected for low redshift cluster lensing, the maximum tangential ellipticity
is only $4\%$, while the majority of the signal is $\sim 1\%$. The major contribution to    
$\chi^2$ comes from a few bins located at the edge of the
images; removing the two worst bins drops $\chi^2$ substantially, 
raising the probability that the ellipticities are consistent within 
the errors to 
$\sim 30\%$. The edges of our images have the largest field distortions and
shallowest coverage and are therefore where we expect the largest
discrepancy between the data sets.  For the blue filters, due to the shallowness of the $\beta$ set and
the fact that the $B$ and $B_J$ filters are not identical, 
there are very few background objects in common ($\sim$
1000 vs. 9000 in the R images).  
The error in $\chi^2$ is dominated by the rms ellipticity, 
giving a value of $\chi^2 = 1.125$  
or  $P(\chi^2|200)=10.9\%$.
This is a very strict test of our measurement and correction
algorithm, showing that our shear measurements are reproducible. The relatively low
formal probabilities may be due to either small residual systematics or slight
underestimates of the measurement error. Across the image, the high correlation between the corrected 
data sets is quite encouraging, with large differences generally 
confined to small areas near the edge of the field.

We performed mass reconstruction on a $60 \times 60$ grid,  
applying a version of the Kaiser and Squires \cite{squ96} algorithm 
separately 
to the blue and red catalogs and combining them by weighting the mass 
at each gridpoint by its S/N. We have chosen to plot the S/N of each mass map pixel
as this gives a direct picture of which mass peaks are significant.  We estimated the 
the noise in each pixel for this map as follows: we rotated each galaxy orientation 
through an arbitrary angle and then computed the resulting mass map; we then repeated
this procedure 100 times and estimated the noise from the variance of these 100 noise 
maps.   As a check of systematics, we produced mass maps for each image with all galaxy
orientations rotated by 45 degrees and found no significant features.  In the absence of
residual systematic effects such a map should be consistent with noise
\citep{ste96,kai94}. As an additional systematics check, we have made shear maps for the
stars in each filter; they are consistent with noise, as expected.

The combined 
convergence maps are shown in Figures 2 and 3 for the 
$\alpha$ and $\beta$ data sets. There is remarkable agreement in morphology of the two maps, 
and the largest mass features appear to be robust.  To quantify this correlation, we calculate Pearson's r between the two mass maps: $$r = \frac{\sum^N (m_{\alpha} - \langle
m_{\alpha} \rangle)(m_{\beta}-\langle m_{\beta} \rangle)}{\sqrt{\sum^N
(m_{\alpha}- \langle m_{\alpha} \rangle)^2}\sqrt{\sum^N (m_{\beta}- \langle
m_{\beta} \rangle)^2}},$$  where $m$ is the value of a mass map pixel and $N$ 
is the total number of gridpoints. $r=-1$ for completely anticorrelated
data, 0 for uncorrelated data, and 1 for completely correlated data.  
We find a value of
$r=0.6$ between the two maps, indicating that they are in fact correlated.  
The formal probability of achieving such a high value of $r$ by chance for 
uncorrelated maps is $\rm{erfc}(|r|\sqrt{N/2})$, negligibly small for  
3600 gridpoints. To further quantify the degree of correlation, we calculated
Pearson's r between the 100 noise maps described above and found these maps gave a value of $-0.009 \pm 0.089$. To determine the correlations between the entire maps, rather than
the correlation introduced by the coincidence of the central peak, we masked out a
box centered on the two maps with a size of $10'$.  When Pearson's r was computed on
the remaining unmasked areas, we still found a value of $r=0.40$. This value remained
fairly constant as we increased the masked region until the masked region's size
approached that of the two images. Inspection of the maps indicates that even 
the relatively low significance ($\sim 3 \sigma$) mass peaks which correspond to a
convergence of $\sim 0.02$ are reproducible.  We quantified this by masking regions
greater than $3 \sigma$ in either map and computing Pearson's r on the unmasked
portions.  The value of the correlation only dropped to $r=0.27$.  We note that r
dropped quickly to zero if we 
masked out all regions of significance lower than $3 \sigma$. 
We have also implemented the mass reconstruction algorithm  of Seitz 
\& Schneider (1998) and find the same correlation between features.

The mass map is strongly peaked around the central D galaxy and generally correlates well
with both the cluster light and ROSAT X-ray flux distributions \citep{jof00}.  At
lower significance, there also appears to be mass associated with the second  bright D galaxy 
in the NW of the image and with cluster galaxies in the N and SE of the central D.

 \section{Conclusion}
We have detected weak lensing at the $0.5-4\%$ level 
in a nearby galaxy cluster using two separate sets
 of images.  Despite differences in depth, seeing, and filters,
the shear maps are consistent within the errors, and the reconstructed 
mass maps are strongly correlated. This reproducibility demonstrates the feasibility of using weak lensing to probe the mass distribution 
in low-redshift clusters. In future work, we will apply these methods 
to all the clusters in our survey and compare the resulting mass 
maps with X-ray, optical, and Sunyaev-Zel'dovich data to obtain 
a more detailed picture of the properties of nearby clusters.

\acknowledgements
This work is supported in part by Chandra Fellowship grant PF8-1003,
awarded through the Chandra Science Center.  The Chandra Science
Center is operated by the Smithsonian Astrophysical Observatory for
NASA under contract NAS8-39073. This research was supported in part 
by the DOE and by NASA grant NAG5-7092 at Fermilab and NSF PECASE grant AST 9703282
at the University of Michigan.



{}
\clearpage
\onecolumn
\begin{figure}
\epsscale{.6}
\plotone{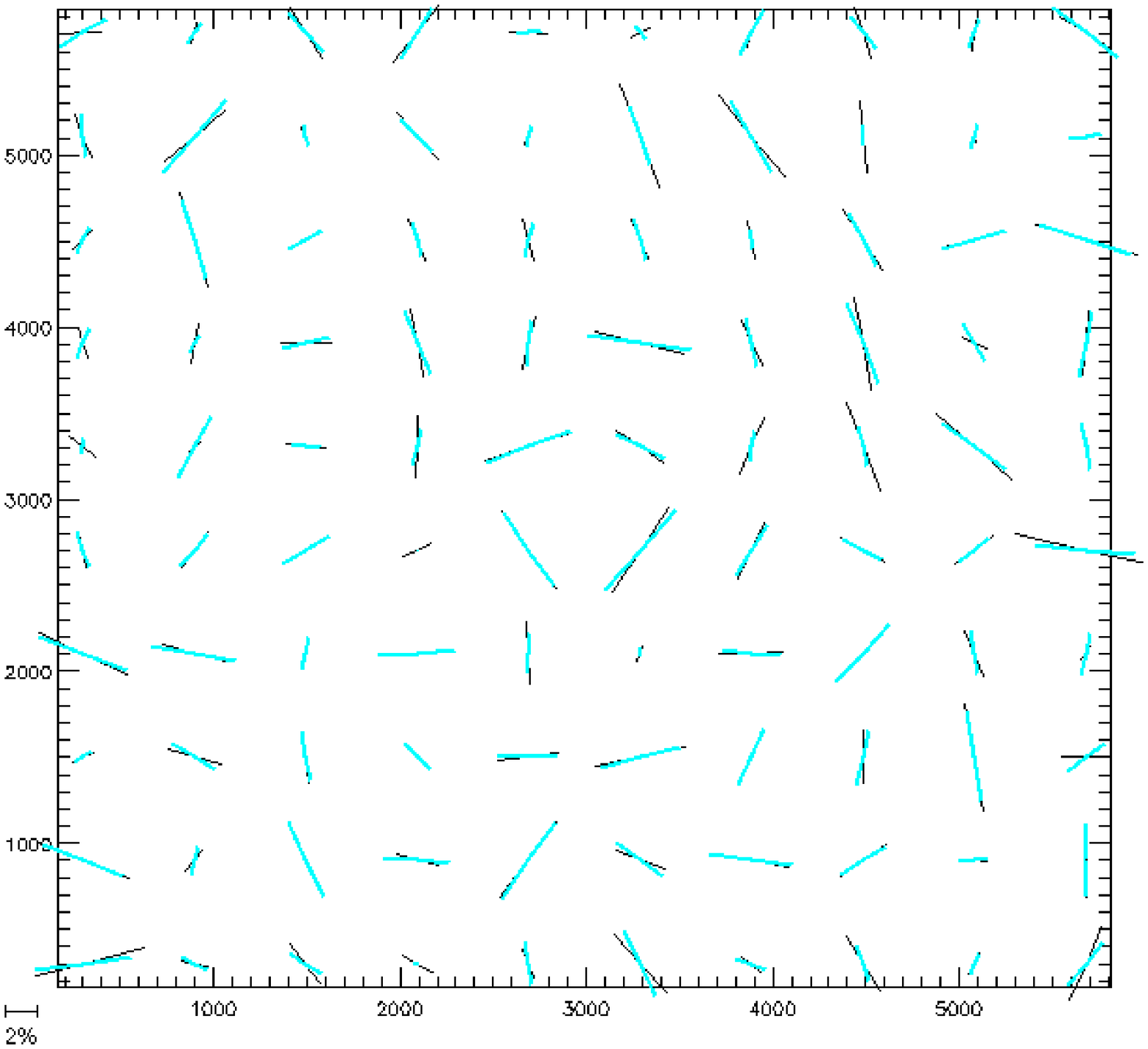}
\plotone{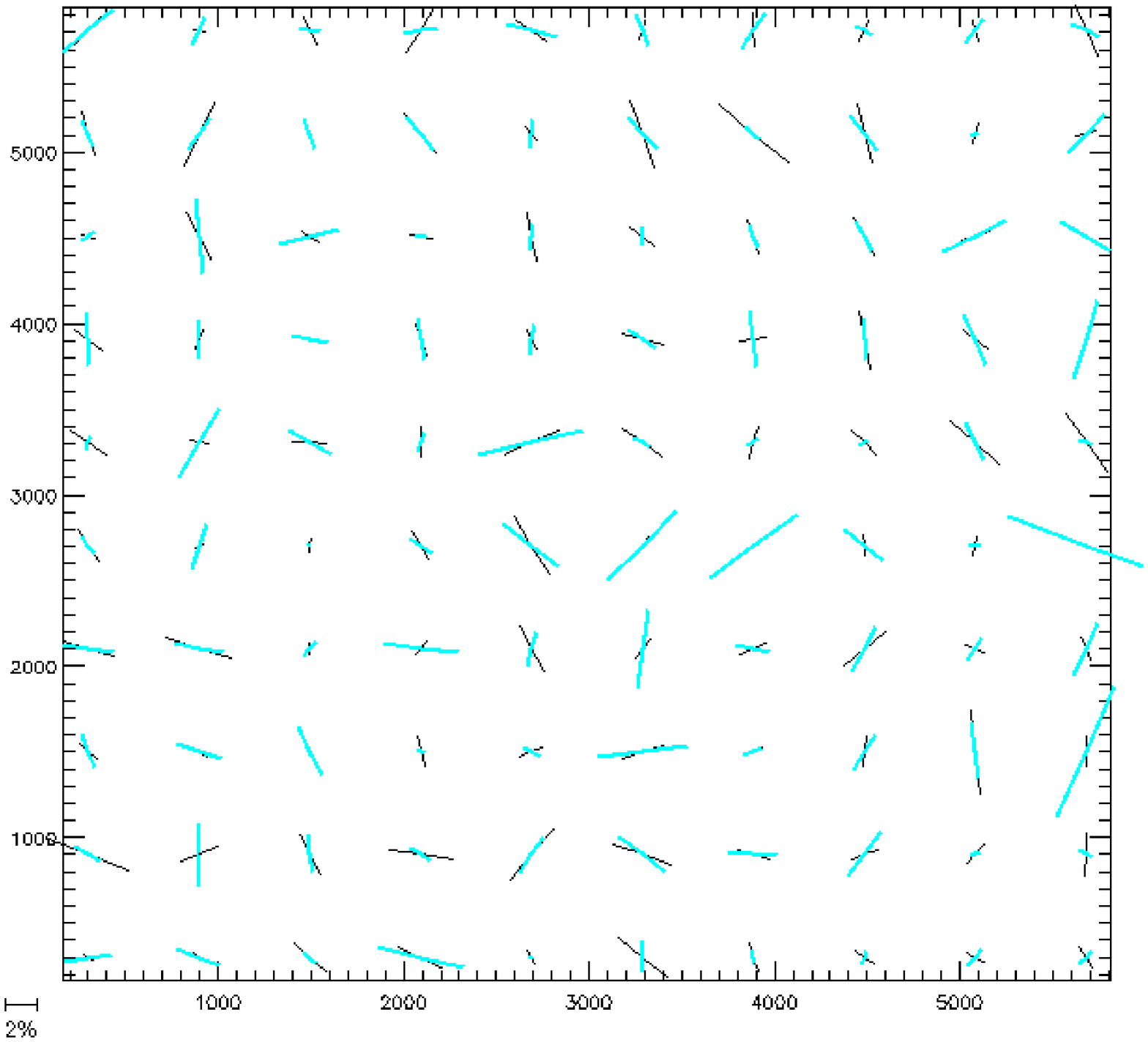}
\figcaption{(a) The binned average ellipticities of background galaxies
appearing in both R band images,  light blue for the $\alpha$ data set and black for
the $\beta$ data set. (b) The same for all background galaxies.  The units are CCD
pixels; an ellipticity of $2 \%$ is shown in the lower left.}
\end{figure}
\begin{figure}
\epsscale{1}
\plotone{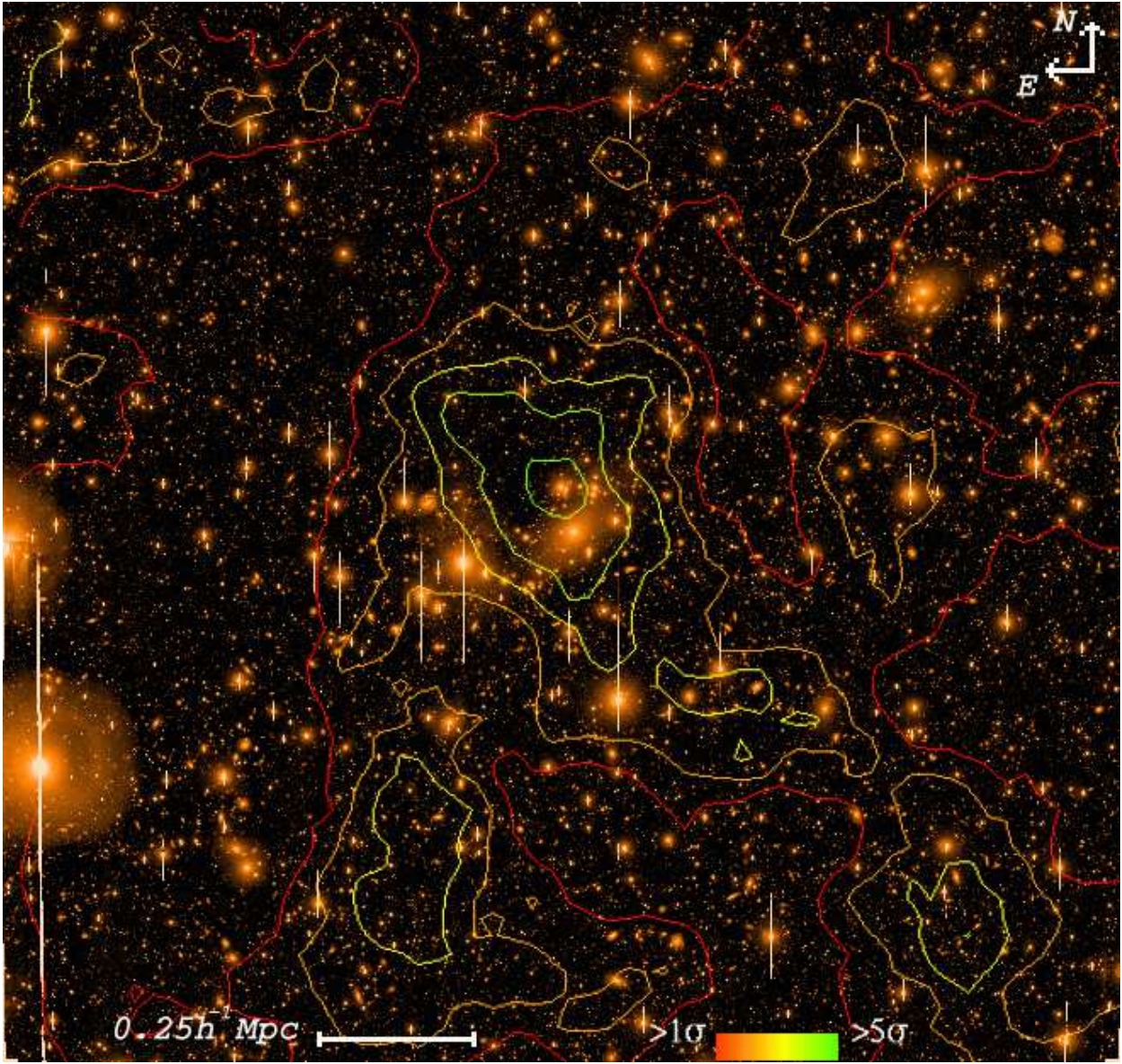}
\caption{Projected mass map of A3667 for set $\alpha$. Contours correspond to a change in signal to noise of unity; only contours $\ge 1
\sigma$ are shown. The mass map is superimposed upon the $\alpha$ $R$ image. The image is $42' \times 42'$.}
\end{figure}
\clearpage
\begin{figure}
\plotone{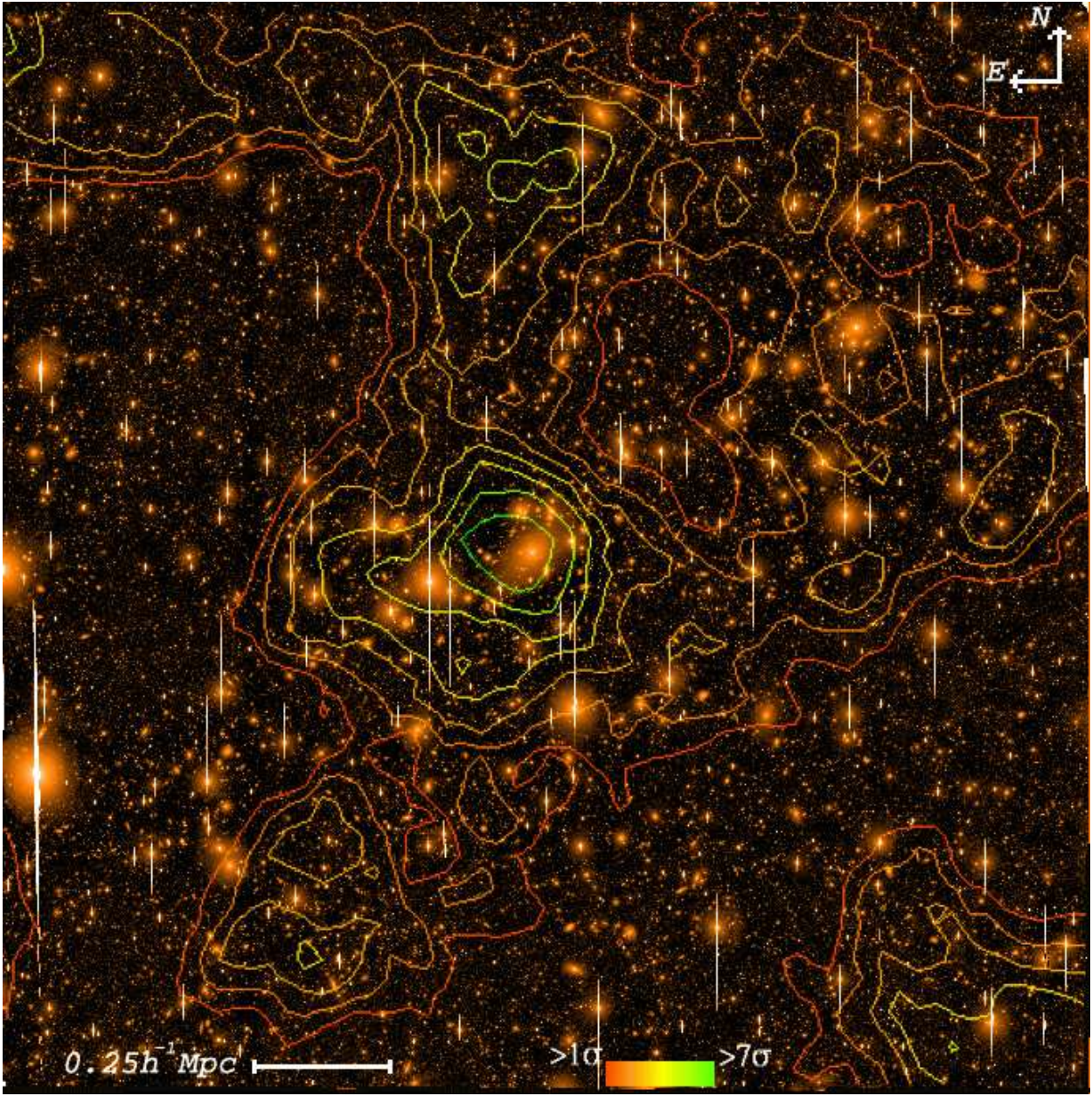}
\figcaption{Projected mass map of A3667 for set $\beta$. The mass map is superimposed upon the $\beta$ $R$ image. The 
image is $44' \times 44'$.}
\end{figure}


\end{document}